\documentclass[12pt,tightenlines,eqsecnum,floats,aps,amsmath,amssymb,nofootinbib,prd,superscriptaddress,showpacs]{revtex4}

\usepackage{setspace}
\usepackage{amsmath,amssymb,amsfonts,amsthm}
\usepackage{graphicx}
\usepackage{enumerate}

\newtheorem{thm}{Theorem}[section]
\newtheorem{prop}[thm]{Proposition}%[section]
\begin{document}

\title{Quantum Einstein-Rosen waves: Coherent states and $n$-point functions}

\author{J. Fernando \surname{Barbero G.}}
\email[]{fbarbero@iem.cfmac.csic.es} \affiliation{Instituto de
Estructura de la Materia, CSIC, Serrano 123, 28006 Madrid, Spain}
\author{I\~naki  \surname{Garay}}
\email[]{igael@iem.cfmac.csic.es} \affiliation{Instituto de
Estructura de la Materia, CSIC, Serrano 123, 28006 Madrid, Spain}
\author{Eduardo J. \surname{S. Villase\~nor}}
\email[]{ejsanche@math.uc3m.es} \affiliation{Instituto Gregorio Mill\'an, Grupo de Modelizaci\'on
y Simulaci\'on Num\'erica, Universidad Carlos III de Madrid, Avda.
de la Universidad 30, 28911 Legan\'es, Spain} \affiliation{Instituto
de Estructura de la Materia, CSIC, Serrano 123, 28006 Madrid, Spain}

\date{July 24, 2008}

\begin{abstract}
We discuss two different types of issues concerning the quantization of Einstein-Rosen  waves. First of all we study in detail the possibility of using the coherent states corresponding to the dynamics of the auxiliary, free Hamiltonian appearing in the description of the model to study the full dynamics of the system. For time periods of arbitrary length we show that this is only possible for states that are close, in a precise mathematical sense, to the vacuum. We do this by comparing the quantum evolutions defined by the auxiliary and physical Hamiltonians on the class of coherent states. In the second part of the paper we study the structure of $n$-point functions. As we will show their detailed behavior differs from the one corresponding to standard perturbative quantum field theories. We take this as a manifestation of the fact that the correct approximation scheme for physically interesting objects in these models does not lead to a power series expansion in the relevant coupling constant but to a more complicated asymptotic behavior.

\end{abstract}


\maketitle

\section{Introduction}{\label{Intro}}

Einstein-Rosen (ER) waves \cite{Einstein} provide a very interesting toy model to discuss several issues relevant for the quantization of general relativity \cite{Kuchar:1971xm, Romano:1995ep, Ashtekar:1996bb, Ashtekar:1996yk, Gambini:1997jj, Dominguez:1999sq, Angulo:2000ad, Varadarajan:1999aa,BarberoG.:2003ye, BarberoG.:2003pz,BarberoG:2004uc, BarberoG.:2004uv, BarberoG.:2005ge, FernandoBarbero:2006gd,Varadarajan:2006af,Cho:2006zz}. The reason behind this is the possibility of exactly describing the dynamics of the system both classically and quantum mechanically. This is true even after coupling some types of matter fields --massless scalars-- to the gravitational degrees of freedom \cite{BarberoG.:2005ge,FernandoBarbero:2006gd}. The main purpose of this paper is to discuss two different issues. The first is related to the problem of finding semiclassical states for the dynamics of the system, the second is to discuss the structure of important physical objects: the $n$-point functions of the model.

A Hamiltonian description of ER-waves shows that the dynamics of this system is rather interesting owing to some unexpected features of the model. Probably the most striking one is that the Hamiltonian is bounded both above and below. This is a direct consequence of the fact that it can be written as a bounded function of the Hamiltonian for a free field theory \cite{Ashtekar:1994ds, Varadarajan:1995hw}. The origin of this free Hamiltonian can be traced back to the asymptotic behavior chosen for the metric at infinity (in the 2+1 dimensional sense explained in \cite{Ashtekar:1994ds}). It is possible to gain a lot of information on the classical and quantum dynamics of the system by taking advantage of this functional dependence mentioned above. For instance, the quantization can be carried out by using a Fock Hilbert space defined by the auxiliary free Hamiltonian. This is so because the spectral theorem allows us to define the physical Hamiltonian  once we construct the auxiliary one in a suitable Hilbert space.

The first issue that we want to discuss here is related to the definition of semiclassical states for the full dynamics and the possibility of using the ones corresponding to the free auxiliary Hamiltonian as an approximate substitute. Coherent states play a very important role for systems of coupled harmonic oscillators --including free field theories that can be readily interpreted as models consisting of an infinite number of them. They display a very interesting behavior because they somehow bridge the gap between states with a purely quantum behavior and classical solutions to the equations of motion. For example, even though they display the characteristic dispersion of position and momentum observables, coherent states are minimal in the sense that the Heisenberg inequalities are saturated. Also the mean values of position and momenta evolve according to the classical equations of motion. For this reason coherent states can be considered as the best semiclassical states for linear systems. In the context of ER-waves, the large quantum gravity effects discovered by A. Ashtekar in \cite{Ashtekar:1996yk} (see also \cite{Gambini:1997jj,Dominguez:1999sq,Angulo:2000ad}) were analyzed by using coherent states. The main result in \cite{Ashtekar:1996yk} is that if one considers a coherent state $\Phi_C$ for the quantum scalar field that describes the local degrees of freedom of an ER-wave and computes the relative uncertainties $\Delta_{\Phi_C} \mathcal{O}/\langle \mathcal{O}\rangle_{\Phi_C}$ of a certain relevant observable $\mathcal{O}$ --that can be interpreted as the quantum counterpart of a metric component of the ER-waves-- one gets huge uncertainties even for coherent states with low (but no-zero) occupation number\footnote{Written in terms of the classical initial data $C$.} $\|C\|^2$ when $C$ is peaked around a not too low value of the energy.  Here, we want to discuss, from a \textit{dynamical} point of view, the usefulness  of coherent states as \textit{bona fide} semiclassical states. As we will see, the class of coherent states can not be considered as semiclassical for the dynamics of the system and hence the behavior discussed in \cite{Ashtekar:1996yk,Gambini:1997jj,Dominguez:1999sq,Angulo:2000ad} is, in fact, rather natural from our point of view. The results of this paper should be considered as complementary to those presented in \cite{Ashtekar:1996yk,Gambini:1997jj,Dominguez:1999sq,Angulo:2000ad}.

The problem of finding semiclassical states for systems different from the harmonic oscillator is a very difficult one whose general solution is not known (in fact, even for such important systems as the hydrogen atom no such states have been found to date). As a consequence of this it is natural to expect that no coherent states --in the traditional sense-- exist for the dynamics of the ER-waves (in fact this is a consequence of a simple exercise that can be carried out for the one-dimensional harmonic oscillator \cite{coher}). What we want to study here is to what extent the coherent states that \textit{do exist} for the auxiliary dynamics can be used to derive meaningful information for quantum ER-waves. To this end, we will compare the states obtained by evolving a coherent state $\Phi_C$ for the free Hamiltonian at a certain time $t_0=0$ both with the auxiliary dynamics $U_0(t)$ and the full physical dynamics\footnote{Notice that we can use the same Hilbert space to describe the auxiliary and the full dynamics because the physical Hamiltonian is a function of a free Hamiltonian.} $U(t)$. We do this by considering $\|U_0(t)\Phi_C-U(t)\Phi_C\|$, and the projection of $U(t)\Phi_C$ on the coherent state $\Phi_{C_t}$ labeled by the classical evolution of the classical initial data $C$. We do this both for small and large values of $t$. As we will see it is possible to quantify the periods of time where free coherent states remain approximately coherent for the full dynamics. We will also show that for large values of $t$ their distance becomes as large as possible (for orthogonal states).

A second set of questions that we want to address concerns the relationship between the exact quantization of ER-waves and standard perturbative approaches. As the system can be exactly solved it is possible to devise efficient approximation schemes to extract physical information about it. We have done this in the past to discuss, for example, issues related to microcausality \cite{BarberoG.:2003ye,BarberoG:2004uc,BarberoG.:2004uv}. The main lesson that we have learnt from this type of analysis is that the asymptotic behavior of physically interesting objects is not captured by simple power series expansions in terms of the relevant coupling constant. Here we give an alternative way to understand this by looking at $n$-point functions. These are the building blocks used in standard perturbative analysis in QFT to obtain the $S$-matrix and discuss such important issues as renormalizability. As we will see the structure of $n$-point functions is such that one cannot expect a simple perturbative series to appear and, hence, the non-standard asymptotic behaviors found in previous works are natural in this setting.

The paper is organized as follows. After this introduction we discuss in section \ref{Canonical reduced phase space} the basic material needed to describe the canonical reduced phase space for the model. The next section \ref{1phs} is devoted to the construction of the one-particle Hilbert space for the auxiliary Hamiltonian. Section \ref{Fock} deals with the Fock quantization and the quantum dynamics of the system. We discuss several issues related to the definition of coherent states for the free auxiliary dynamics of the model in section \ref{coherent}. We also study the asymptotics, both for small and large times, of some functions that measure the deviation of the states evolved with the full dynamics with respect to the states obtained by evolving the same quantum initial data with the free, auxiliary, dynamics. The main point discussed in section \ref{Npoint} is the structure of the $n$-point functions. They can be exactly written in closed form.  At every relevant order $n$ we uncover a mixing property that characterizes their structure and shows that we are dealing with an interacting theory despite the fact that the main building block in its construction is a free model. We end with the conclusions and an appendix that summarizes some technical results concerning the asymptotic expansions that appear in the main body of the paper.

\section{Canonical reduced phase space}{\label{Canonical reduced phase space}}

Einstein-Rosen waves describe vacuum solutions to the Einstein equations\footnote{Here and in the following we use Penrose's abstract index notation.} $R^{{\scriptscriptstyle(4)}}_{ab}=0$    for a symmetry reduction of general relativity consisting of space times of the form $(\mathbb{R}^4,g_{ab}^{{\scriptscriptstyle(4)}})$ with four-dimensional metrics $g_{ab}^{{\scriptscriptstyle(4)}}$ having two hypersurface orthogonal, commuting, spatial Killing vector fields. The isometry group of these space-times is  $\mathbb{R}\times U(1)$ and the metrics are regular at a symmetry axis. If we use a single global coordinate chart $(x,y,z,t)$ on $\mathbb{R}^4$ it is possible to take these Killing fields as $(\partial/\partial z)^a$ and $(\partial/\partial \sigma)^a:=x(\partial/\partial y)^a-y(\partial/\partial x)^a$. The cylindrical coordinates naturally associated to the previous cartesian coordinates allow us to write the four-dimensional metric in the form
\begin{eqnarray*}
g_{ab}^{{\scriptscriptstyle(4)}}=e^{\gamma-\phi}\big[-e^{-\gamma_\infty}(\mathrm{d}t)_a(\mathrm{d}t)_b
+(\mathrm{d}r)_a(\mathrm{d}r)_b\big]
+r^2e^{-\phi}(\mathrm{d}\sigma)_a(\mathrm{d}\sigma)_b
+e^{\phi}(\mathrm{d}z)_a(\mathrm{d}z)_b\,
\end{eqnarray*}
where  $-\infty <t<\infty$, $0<r=\sqrt{x^2+y^2}$, $0<\sigma<2\pi$, $-\infty <z<\infty$ and the symmetry axis lies at $r=0$. In the previous expression $\gamma(t,r)=\gamma(t,\sqrt{x^2+y^2})$ and $\phi(t,r)=\phi(t,\sqrt{x^2+y^2})$ are smooth functions\footnote{These conditions guarantee the smoothness of the axially symmetric scalar function at $r=0$. In the following when we talk about smoothness in the axis we will refer to this condition.} of $(t,x,y)\in \mathbb{R}^3$, and $\gamma_\infty(t):=\lim_{r\rightarrow \infty}\gamma(t,r)$. Notice that $e^\phi$ is the norm of the translational Killing field $(\partial/\partial z)^a$. The time coordinate $t$ is chosen in such a way that the vector field $(\partial/\partial t)^a$ is an asymptotic (for large $r$) unit Killing vector field for all the 2+1 dimensional metrics of the form
\begin{eqnarray}
g_{ab}=e^\phi\big(g_{ab}^{{\scriptscriptstyle(4)}}-e^\phi(dz)_a(dz)_b\big)=
-e^{\gamma-\gamma_\infty}(\mathrm{d}t)_a(\mathrm{d}t)_b+e^{\gamma}(\mathrm{d}r)_a(\mathrm{d}r)_b
+r^2(\mathrm{d}\sigma)_a(\mathrm{d}\sigma)_b\label{g}
\end{eqnarray}
defined on the space of orbits of the translational Killing vector field. This choice makes sense \cite{Ashtekar:1996bb} because, under certain fall-off conditions for the fields $\gamma$ and $\phi$,  the Einstein equations imply that $\gamma_\infty$ is time independent and non-negative. The value of
\begin{eqnarray}
2\pi(1-e^{-\gamma_\infty/2})\in[0,2\pi)\label{deficit}
\end{eqnarray}
represents the deficit angle of the asymptotically conical, $2+1$ dimensional metrics, belonging to the class defined by (\ref{g}).  We want to remark at this point that when $\gamma_\infty=0$ this deficit angle vanishes and, hence, the asymptotic behavior is exactly given by the auxiliary Minkowskian background metric
\begin{eqnarray}
\eta_{ab}=-(\mathrm{d}t)_a(\mathrm{d}t)_b
+(\mathrm{d}r)_a(\mathrm{d}r)_b
+r^2(\mathrm{d}\sigma)_a(\mathrm{d}\sigma)_b\,.\label{eta}
\end{eqnarray}
This background metric will  play an important role in the following.

\bigskip

As we have mentioned above, the Einstein field equations $R^{{\scriptscriptstyle(4)}}_{ab}=0$ force $\gamma_\infty$ to be  constant in $t$ whereas the function $\gamma$ can be obtained in terms of $\phi$ (see, for example, \cite{Ashtekar:1996bb, FernandoBarbero:2006gd}). In particular, once we fix some Cauchy data $(Q,P)$  for $\phi$ at some initial time $t=t_0$ (say, $t_0=0$), we can compute the quantity
$$
\gamma_\infty=\frac{1}{2}\int_0^\infty\Big(P^2(r) +Q'^2(r) \Big) r\mathrm{d}r\,,
$$
--that will be seen to be a constant of motion under the dynamics defined by the Hamiltonian given below-- solve for $\phi(t,r)$ as a solution to the Einstein equations with initial data
$$
\phi(0,r)=Q(r)\,,\quad e^{\gamma_\infty/2}\dot{\phi}(0,r)=P(r)\,\quad r\in[0,\infty)\,
$$
and, finally, obtain
\begin{eqnarray*}
\gamma(t,r)=\frac{1}{2}\int_0^r\Big(e^{\gamma_\infty} \dot{\phi}^2(t,s) +\phi'^2(t,s) \Big) s\mathrm{d}s\,.
\end{eqnarray*}
As we can see, the local physical degrees of freedom of the ER waves are described by a scalar field $\phi$.

\bigskip

The dynamics induced by the Einstein equations on the field $\phi$ that we have just described admits a well-known Hamiltonian formulation \cite{Ashtekar:1996bb} for which the Cauchy surfaces are the level surfaces of the asymptotic Minkowskian time coordinate $t$. In order to describe it let us first introduce $\mathcal{C}\subset C^\infty(\mathbb{R}^2)$, the linear space whose points are smooth real functions on $\mathbb{R}^2$ with rapid decay that depend on $(x,y)\in\mathbb{R}^2$ through $r=\sqrt{x^2+y^2}\in[0,\infty)$. The asymptotic conditions for the Cauchy data $Q(r)$ and $P(r)$ can be relaxed both at $r=0$ and $r\rightarrow\infty$. This is important for the classical viewpoint but is irrelevant for the Fock quantization considered in this paper in the sense that the Fock space of quantum states turns out to be insensitive to the detailed choice of these asymptotic conditions.
The canonical reduced phase space of the Einstein-Rosen waves  $\Upsilon=(\mathcal{P},\omega)$, with points generically denoted as $(Q,P)\in \Upsilon$, is defined by endowing
$\mathcal{P}=\mathcal{C}\times \mathcal{C}$
with the standard (weakly) symplectic structure
\begin{equation}\label{omega}
\omega((Q_1,P_1),(Q_2,P_2)):=\int_0^\infty
\Big(Q_2(r)P_1(r)-Q_1(r)P_2(r)\Big)\,r\mathrm{d}r\,.
\end{equation}
The description of the classical dynamics in $\Upsilon$ is done in the form of an autonomous Hamiltonian system
$(\Upsilon,\omega,h)$ with a Hamiltonian $h:\Upsilon\rightarrow\mathbb{R}$ that is defined in terms of the quadratic (free) auxiliary Hamiltonian\footnote{We use units such that $c = \hbar = 8G_3 = 1$, where $G_3$
denotes the effective Newton constant per unit length in the direction of the symmetry axis.}
\begin{equation*}
h_0(Q,P):=\gamma_\infty(Q,P)=\frac{1}{2}\int_0^\infty
\Big(P^2(r)+Q'^2(r)\Big)\,r\mathrm{d}r
\end{equation*}
through a non-polynomial map
\begin{equation}
h(Q,P)=2-2\exp\Big(-\frac{1}{2}h_0(Q,P)\Big).\label{full_hamiltonian}
\end{equation}
Notice that, in view of (\ref{deficit}), $h(Q,P)$ can be interpreted (up to a $\pi$ factor) as the deficit angle of the metrics (\ref{g}). The Hamilton equations derived from this non-quadratic Hamiltonian are
\begin{eqnarray*}
\dot{Q}&=&e^{-h_0(Q,P)/2}P\\
\dot{P}&=&e^{-h_0(Q,P)/2} \Delta Q
\end{eqnarray*}
where $\Delta$ denotes the Laplacian $\Delta:\mathcal{C}\rightarrow \mathcal{C}$ acting on axially symmetric functions as
$$
(\Delta F)(r):=F''(r)+\frac{F'(r)}{r}\,
$$
Notice that the $1/r$ term in the previous expression originates in the axial symmetry of the system.

\section{The one-particle Hilbert space adapted to the asymptotic structure}{\label{1phs}}

This section is devoted to the construction of the one-particle Hilbert space that we will later use to build the Fock space where the quantization of this system will take place. We start by pointing out that the Laplacian operator introduced above can be extended to a densely defined operator $\Theta=-\Delta$ on $L^2([0,\infty),r\mathrm{d}r)$. The operators $\Theta$ and $\sqrt{\Theta}$ are self-adjoint and non-negative. As usual\footnote{Here we closely follow the ideas developed in \cite{Wald}.} it can be employed to define a complex structure $J:\Upsilon\rightarrow \Upsilon$ on the canonical phase space according to\footnote{In order to make sense of $J$ it is necessary to restrict the domain of $\Theta\geq0$ so that $1/\sqrt{\Theta}$ is well defined. To this end it suffices to consider functions $F(x,y)$ such that their Fourier transform $f(w_1,w_2)$ vanishes in a neighborhood of zero. This guarantees that $\sqrt{w_1^2+w_2^2}f(w_1,w_2)$ and $f(w_1,w_2)/\sqrt{w_1^2+w_2^2}$ are of rapid decay and smooth, even at $(w_1,w_2)=(0,0)$, when  $f(w_1,w_2)$ is chosen smooth and of rapid decay.   We will implicitly use this domain when needed. Notice however that there are  many functions that can be used to describe physical situations, for example the gaussian $e^{-x^2-y^2}$,  that do not satisfy this restriction.}
\begin{eqnarray}
J\left(\begin{array}{c}Q\\P\end{array}\right):=\left(
\begin{array}{cc} 0 &  -1/\sqrt{\Theta}\\
  \sqrt{\Theta} & 0\end{array}
\right) \left(\begin{array}{c}Q\\P\end{array}\right) \,.\label{J}
\end{eqnarray}
This complex structure is the restriction to the axisymmetric case of the standard complex structure adapted to the  Poincar\'e symmetry of the background metric (\ref{eta}).  It can be used to construct a complex vector space $\Upsilon_J$ whose points are \textit{exactly the same} as the points of $\Upsilon$ and the multiplication by complex numbers\footnote{Here  $x$, $y\in \mathbb{R}$. As usual  $i=\sqrt{-1}$ is the imaginary  unit.} $x+iy\in\mathbb{C}$ is defined by
$$
(x+iy)(Q,P):=x(Q,P)+yJ(Q,P)\,.
$$
It is possible to combine now $\omega$ and $J$ to define a positive definite sesquilinear form
\begin{eqnarray*}
& & \langle\cdot\,,\cdot \rangle_J:\Upsilon_J\times\Upsilon_J\rightarrow \mathbb{C}\,,\\
& & \langle(Q_1,P_1),(Q_2,P_2) \rangle_J=\frac{1}{2}\omega
(J(Q_1,P_1),(Q_2,P_2))-\frac{i}{2}\omega((Q_1,P_1),(Q_2,P_2))\,.
\end{eqnarray*}
providing us with a scalar product on $\Upsilon_J$. The one-particle Hilbert space $\mathcal{H}_J$ of the ER waves is the Cauchy completion of  $(\Upsilon_J,\langle \cdot\,,\cdot \rangle_J)$.

\bigskip

There is another useful construction of the one-particle Hilbert space that uses  $\Upsilon_\mathbb{C}$ --the $\mathbb{C}$-vector space obtained from $\Upsilon$ by considering complex functions in $\mathcal{C}_{\mathbb{C}}\subset C^\infty([0,\infty),\mathbb{C})$ with the standard multiplication by complex scalars-- as the starting point. In order to see this we notice that the complex structure (\ref{J}) can be diagonalized in $\Upsilon_\mathbb{C}$. In fact, the vectors
$$
(C,\mp i \sqrt{\Theta}C)
\in\Upsilon_\mathbb{C}\,,\quad C\in\mathcal{C}_\mathbb{C}\,,
$$
are eigenvectors of $J$ corresponding to the eigenvalues $\pm i$. Hence we can write  $\Upsilon_\mathbb{C}$ as the direct sum $\Upsilon_\mathbb{C}=\Upsilon_+\oplus \Upsilon_-$
where
$$
\Upsilon_\pm:=\{(C,\mp i \sqrt{\Theta}C)\in
\Upsilon_\mathbb{C}\,|\, C\in \mathcal{C}_\mathbb{C}\}\,.
$$
It is clear that $\Upsilon_+\cap \Upsilon_-=\{0\}$ and $\bar{\Upsilon}_+=\Upsilon_-$ where
$$\overline{(C,\mp i \Theta^{1/2}C)}:=(\overline{C},\pm i \Theta^{1/2}\overline{C}).$$
If we take a point  $(Q,P)\in \Upsilon$  there exists a unique  $C\in \mathcal{C}_\mathbb{C}$, given by
$$
C=\frac{1}{2}\bigg(Q+i\Theta^{-\frac{1}{2}} P\bigg)\,,
$$
such that
\begin{equation}
(Q,P)=(C,-i\sqrt{\Theta}C)+\overline{(C,-i\sqrt{\Theta}C)}\,.\label{def_C}
\end{equation}
From equation (\ref{def_C}) it is easy to see that given the first component $C=\frac{1}{2}(Q+i\Theta^{-\frac{1}{2}} P)$ of $(C,-i\sqrt{\Theta}C)\in \Upsilon_+$; the other can be then computed without any ambiguity. Hence, the one-particle Hilbert space $\mathcal{H}_J$ can be equally well described in terms of the complex functions $C$ by using the following identification\footnote{At this point, $\mathcal{H}$ is just a linear space of complex functions $C(r)$ that will become a Hilbert space once a scalar is introduced.} $\kappa:\mathcal{H}_J\rightarrow \mathcal{H}$
\begin{eqnarray}
C&=&\kappa(Q,P):=\frac{1}{2}\bigg(Q+i\Theta^{-\frac{1}{2}} P\bigg)\,,\label{kappa1}
\\
(Q,P)&=&\kappa^{-1}C=(C+\overline{C},-i\sqrt{\Theta}(C-\overline{C}))\,.\label{kappa2}
\end{eqnarray}
The map $\kappa$ is adapted to the complex form $J$ in the sense that if $\kappa(Q,P)=C$ then $\kappa {\small\circ} J(Q,P)=iC$. According to the previous discussion we can build a Hilbert space
\begin{eqnarray*}
\mathcal{H}=\{C\,:\,\|C\|^2=\langle C,C\rangle<\infty\},
\end{eqnarray*}
defined with the help of the scalar product
\begin{eqnarray*}
\langle
C_1,C_2\rangle&:=&\frac{1}{2}\omega(J\kappa^{-1}C_1,\kappa^{-1}C_2)-\frac{i}{2}\omega(\kappa^{-1}C_1,\kappa^{-1}C_2),
\end{eqnarray*}
that is equivalent to the one-particle Hilbert space $\mathcal{H}_J$. The complex structure in $\mathcal{H}$ is diagonal ($J C=iC$ for all $C\in \mathcal{H}$). At variance with the situation concerning the spaces $\Upsilon_{\pm}$ (for which $\bar{\Upsilon}_+=\Upsilon_-$), if we work with $\mathcal{H}$ the operator $\overline{\phantom{C}}$ is a conjugation in $\mathcal{H}$, that is, $\mathcal{H}$ is an antilinear map from $\mathcal{H}$ to $\mathcal{H}$ satisfying $\overline{\overline{C}}=C$.

\bigskip

There are several mathematical structures that are easier to handle in $\mathcal{H}$ than in $\mathcal{H}_J$. For example, the classical Hamiltonian that describes the dynamics of the Einstein-Rosen waves can be written now in terms of the scalar product and the operator $\Theta$ given above by noticing that
\begin{eqnarray*}
h_0(\kappa^{-1}C) &=&\|\Theta^{\frac{1}{4}}C\|^2=\langle
C,\sqrt{\Theta} C\rangle \\
h(\kappa^{-1}C)&=&2-2\exp(-\langle
C,\sqrt{\Theta} C\rangle/2)\,.
\end{eqnarray*}
It is interesting to point out here that the scalar product $\langle \cdot\,,\cdot \rangle$ on $\mathcal{H}$ can be written in terms of the usual $L^2([0,\infty),r\mathrm{d}r)$ product
$$
\langle C_1,C_2\rangle_{L^2}:=\int_0^\infty \overline{C_1(r)}
C_2(r)\,r\mathrm{d}r
$$
as  $\langle C_1,C_2\rangle =2\langle C_1,\sqrt{\Theta}\,C_2\rangle_{L^2}$. This allows to show that $\Theta$ is self-adjoint in $\mathcal{H}$. This can be seen by using the mode decomposition introduced at the end of subsection \ref{modes} that allows us to write $\Theta$ as a multiplication operator.

\subsection{Classical dynamics in the one-particle Hilbert space}{\label{classdyn}}

Let us consider now the dynamics of the scalar field defined by the auxiliary free Hamiltonian
$$
h_0(Q,P)=\frac{1}{2}\int_0^\infty \Big(P^2(r)+Q'^2(r)\Big)\,r\mathrm{d}r=\langle C,\sqrt{\Theta} C\rangle\,\quad (\textrm{with}\,\, C=\kappa(Q,P))$$
in the one-particle Hilbert space $\mathcal{H}$. To this end we first write the free (linear) Hamilton equations in terms of the fields $C$,
\begin{equation}
\dot C_t=-i\sqrt{\Theta} C_t\,,\quad\textrm{ with initial data }C_0=C\,\textrm{ at }t=0.\label{free_equation}
\end{equation}
The self-adjointness of $\Theta$ in $\mathcal{H}$ allows us to write the general solution $C_t^0$ to the equation (\ref{free_equation}) as  the action of the unitary operator $\exp(-it\sqrt{\Theta})$ on the initial data $C$, i.e.
\begin{eqnarray*}
C_t^0=\exp(-it\sqrt{\Theta})C\,.
\end{eqnarray*}
If $C^0_{\alpha t}=\exp(-it\sqrt{\Theta})C_\alpha$ denotes the evolution from $t=0$ to an arbitrary time $t$ of some initial data that we label as $C_\alpha$, the unitary character of the classical evolution in $\mathcal{H}$ implies that
\begin{eqnarray*}
\langle C^0_{1t},C^0_{2t}\rangle
=\langle C_{1},C_{2}\rangle\quad\textrm{ for all }t\in \mathbb{R}\,.
\end{eqnarray*}
In particular the norms of the states remain constant in time.

\bigskip

The non-linear dynamics defined by (\ref{full_hamiltonian}) on the reduced phase space of the Einstein-Rosen waves  can  be also described in the one-particle Hilbert space $\mathcal{H}$. The solution $C_t$ at time $t$ to the field equations with initial data $C$ at $t=0$ can be written now as
\begin{eqnarray}
C_t=\exp\left(-ite^{-h_0(\kappa^{-1}C)/2}\sqrt{\Theta}\right) C\,.\label{full_solution}
\end{eqnarray}
Notice that $C_t$ depends on the initial data $C$ in a non-linear way. Hence,  if we denote by  $C_{1 t}$, $C_{2t}$, and $C_{(1+2)\, t}$ the solutions corresponding to the initial data $C_1$, $C_2$, and $C_1+C_2$ it is clear that
$C_{(1+2)\, t} \neq C_{1t}+C_{2t}$.   From (\ref{full_solution}) it is also evident that the evolution does not preserve the scalar product on $\mathcal{H}$ i.e.
\begin{eqnarray}
\langle C_{1t},C_{2t}\rangle \neq \langle C_{1},C_{2}\rangle\,.\label{C_1C_2}
\end{eqnarray}
However, it is easy to see from  (\ref{full_solution}) that
\begin{eqnarray*}
\|C_t\|=\|C\|\quad\textrm{ for all }t\in \mathbb{R}\,.
\end{eqnarray*}
This is not in conflict with (\ref{C_1C_2}) because $C_t$ does not depend linearly on $C$.

\subsection{Mode decomposition}{\label{modes}}

In the following we will find it convenient to work with a suitable mode decomposition adapted to the axial symmetry of our system. Given any $F\in \mathcal{C}$ or $\mathcal{C}_\mathbb{C}$, we will use the following Fourier integral representation\footnote{Here $J_0$ denotes the 0th order Bessel function of the first kind.}
$$
F(r)=\frac{1}{\sqrt{2}}\int_0^\infty f(w)J_0(w r)\,
\mathrm{d}w\,
$$
related to the two-dimensional Fourier transform
\begin{eqnarray*}
(\mathcal{F}F)(w)=(\mathcal{F}F)(\sqrt{\smash[b]{w_1^2+w_2^2}}\,)=\frac{1}{2\pi}\int_{\mathbb{R}^2}F(\sqrt{x^2+y^2})
\,e^{-i(xw_1+yw_2)}\,\mathrm{d}w_1\mathrm{d}w_2
\end{eqnarray*}
according to $f(w)=\sqrt{2}w (\mathcal{F}F)(w)$.
The functions $f(w)/w$, where $w=\sqrt{w_1^2+w_2^2}$, belong to the class of $C^{\infty}(\mathbb{R}^2)$ in the two real variables $(\omega_1,\omega_2)$ with rapid decay. Notice that with these conventions $f(0)=0$. The  action of the operator $\sqrt{\Theta}$ can be written in a nice way in this Fourier representation
$$
(\sqrt{\Theta}F)(r)=\frac{1}{\sqrt{2}}\int_0^\infty wf(w)J_0(w r)\,
\mathrm{d}w\,.
$$
The mapping $\kappa$ connecting the $(Q,P)$ and $C$ descriptions of the one-particle Hilbert space also has a simple expression in this representation. In fact, given
\begin{eqnarray*}
Q(r)&=&\frac{1}{\sqrt{2}}\int_0^\infty q(w)J_0(w r)\,
\mathrm{d}w\\
P(r)&=&\frac{1}{\sqrt{2}}\int_0^\infty p(w)J_0(w r)\,
\mathrm{d}w
\end{eqnarray*}
then $C=\kappa(Q,P)$ (we will say that $C$ and $(Q,P)$ are $\kappa$-related) if it can be written as
\begin{eqnarray*}
C(r)&=&\frac{1}{\sqrt{2}}\int_0^\infty c(w)J_0(w r)\,
\mathrm{d}w\,,
\end{eqnarray*}
with
$$
c(w)=\frac{1}{2}\left(q(w)+\frac{ip(w)}{w}\right)\,.
$$
Finally the scalar product of the one-particle Hilbert space $\mathcal{H}$ simplifies to
\begin{eqnarray*}
\langle
C_1,C_2\rangle&=&\int_0^\infty
\overline{c_1(w)}c_2(w)\,\mathrm{d}w\,,\quad \|C\|^2 =\int_0^\infty |c(w)|^2\,\mathrm{d}w\,,
\end{eqnarray*}
and also the free Hamiltonian can be written in the simple form
$$
h_0(\kappa^{-1}C) =\int_0^\infty
w\,|c(w)|^2\,\mathrm{d}w\,.
$$

\bigskip

\section{Fock quantization and quantum dynamics}{\label{Fock}}

The one-particle Hilbert space $\mathcal{H}$ allows us to construct the Hilbert space $\mathcal{F}_s(\mathcal{H})$ used in standard approaches to the quantization of Einstein-Rosen waves as the symmetric Fock space
$$\mathcal{F}_s(\mathcal{H})=\bigoplus_{n=0}^\infty \mathcal{H}^{\otimes_s n}\,
\quad\textrm{ with}\quad  \mathcal{H}^0:=\mathbb{C},$$
where $ \mathcal{H}^{\otimes_s n}$ denotes the symmetrized tensor product of $n$ copies of $\mathcal{H}$. We will write the inner product in $\mathcal{F}_s(\mathcal{H})$ in the form $\langle \cdot  \,|\,\cdot \rangle$.

\bigskip

Following the rules of second quantization \cite{reedsimon} we extend certain operators from $\mathcal{H}$ to the symmetric Fock space $\mathcal{F}_s(\mathcal{H})$. If we are given a unitary operator $\exp(iA):\mathcal{H}\rightarrow\mathcal{H}$ written in terms of a self-adjoint operator $A:\mathcal{D}(A)\subset \mathcal{H}\rightarrow \mathcal{H}$, with domain $\mathcal{D}(A)$, we can promote them to the Fock space $\mathcal{F}_s(\mathcal{H})$. In particular, there exists
a self-adjoint operator
$$\mathrm{d}\Gamma(A):\mathcal{D}(\mathrm{d}\Gamma(A))\subset\mathcal{F}_s(\mathcal{H})\rightarrow \mathcal{F}_s(\mathcal{H})$$
and  a unitary operator $\Gamma(i\exp(A))$ such that
$$
\Gamma(i\exp(A))=\exp(i\mathrm{d}\Gamma(A)):\mathcal{F}_s(\mathcal{H})\rightarrow \mathcal{F}_s(\mathcal{H})\,.
$$
The operator $\mathrm{d}\Gamma(A)$ is called the \textit{second quantization of $A$} and is defined by
\begin{eqnarray*}
\mathrm{d}\Gamma(A):=\bigoplus_{n=0}^\infty A^{(n)}
\end{eqnarray*}
where
\begin{eqnarray*}
A^{(0)}&:=0\,&\\
A^{(n)}&:=&A\otimes I\otimes \cdots \otimes I+I\otimes A\otimes
\cdots \otimes I+\cdots +I\otimes I\otimes \cdots \otimes A\,,
\end{eqnarray*}
and $I$ denotes the identity operator on $\mathcal{H}$. In particular we can use this procedure to construct the free auxiliary Hamiltonian.  This is defined in the one-particle Hilbert space in terms of $\sqrt{\Theta}$ according to
$$
h_0(Q,P)=h_0(\kappa^{-1}C)=\langle C,\sqrt{\Theta}C\rangle\,.
$$
Also, the classical evolution in $\mathcal{H}$  is described in terms of the unitary operator $\exp(-it\sqrt{\Theta})$. Hence, the second quantization of $\exp(-it\sqrt{\Theta})$ and $\sqrt{\Theta}$ will give us the free quantum unitary evolution and the free quantum Hamiltonian. Explicitly, the auxiliary free  Hamiltonian\footnote{We use a lowercase $h_0$ to denote the Hamiltonian quadratic form  in the one-particle Hilbert space and $H_0$ for the free quantum Hamiltonian operator in the Fock space.} $H_0$ is defined on $\mathcal{F}_s(\mathcal{H})$ as
\begin{eqnarray*}
H_0:=\mathrm{d}\Gamma(\sqrt{\Theta})\,,
\end{eqnarray*}
and the quantum evolution operator is given by
\begin{eqnarray}
U_0(t):=\Gamma\Big(\exp(-it\sqrt{\Theta})\Big)
=\exp\Big(-it\,\mathrm{d}\Gamma(\sqrt{\Theta})\Big)=\exp(-itH_0)\,.\label{free_evol}
\end{eqnarray}
Notice that, given $C^{\otimes n}\in\mathcal{H}^{\otimes_s n}$ the action of the free Hamiltonian $H_0$ can be read from the formula
$$
(H_0C^{\otimes n})(r_1,\ldots,r_n)=\frac{1}{2^{n/2}}\int_{[0,\infty)^n}\Big(\sum_{i=1}^nw_i\Big)\prod_{j=1}^nc(w_j)J_0(w_j r_j)\,\mathrm{d}w_j.
$$
Furthermore, notice that $H_0|_{\mathcal{H}}=\sqrt{\Theta}$ and hence, if $C\in \mathcal{H}\subset \mathcal{F}_s(\mathcal{H})$ belongs to the domain of $H_0$, we have\footnote{In the following we use $h_0(C)$ to refer to $h_0(\kappa^{-1}C)$.}
$$
\langle C\,|\, H_0 C\rangle=\int_0^\infty w|c(w)|^2\mathrm{d}w=h_0(C).
$$

\bigskip

In order to make sense of the quantum counterpart of the \textit{full} (physical) classical Hamiltonian (\ref{full_hamiltonian}) we make use of the quantum free Hamiltonian $H_0$ and the functional relation between  the free and physical classical Hamiltonians $h_0$ and $h$. In particular
$$
h(Q,P)=E(h_0(Q,P)):=2-2\exp(-h_0(Q,P)/2)
$$
where $E:[0,\infty)\rightarrow [0,2)$ is the function
\begin{eqnarray}
E(x):=2-2\exp(-x/2)\,.\label{E}
\end{eqnarray}
The spectral theorems then guarantee that the operator
$$
H:=E(H_0)
$$
is a well defined self-adjoint operator on $\mathcal{F}_s(\mathcal{H})$. It is important to notice  that  $H$ \textit{is not} the second quantization of any self-adjoint operator on $\mathcal{H}$. In particular, in spite of the fact that the restriction of $H$ to the one-particle Hilbert space satisfies $H|_{\mathcal{H}}=E(\sqrt{\Theta})$,  the quantum Hamiltonian  $H\neq \mathrm{d}\Gamma(E(\sqrt{\Theta}))$. Hence the unitary operator evolution
$$
U(t)=\exp(-itH)
$$
generated by $H$ \textit{is not} the second quantization of any unitary operator on the one-particle Hilbert space. This is not a surprise because, as we have discussed in section \ref{Canonical reduced phase space}, the full classical dynamics is not even described by a \textit{linear} operator in $\mathcal{H}$. Finally, it is important to point out that the $n$-particle subspaces $\mathcal{H}^{\otimes_s n}$ of the Fock space are stable under the quantum evolution generated by $H$. At first sight this might seem striking because classical ER-waves are not stationary space-times and, in principle, one would expect particle creation effects. However, the asymptotic conditions \cite{Ashtekar:1996bb} used to derive the Hamiltonian formulation discussed in section \ref{Canonical reduced phase space} restrict the class of ER-waves considered here to those metrics that are asymptotically Minkowskian in its $2+1$ formulation (\ref{g}). In this context it is possible to use the preferred Fock quantization associated to the Minkowskian metric (\ref{eta}) for which the particle creation effects are absent.

\section{Coherent states}{\label{coherent}}

As we have discussed  in section \ref{Canonical reduced phase space}, a vector in the one-particle  Hilbert space $C\in \mathcal{H}$ can be thought of, through the identification (\ref{kappa1})-(\ref{kappa2}), as the Cauchy data $(Q,P)$  at a given time for the scalar field that describes the degrees of freedom of an ER-wave. It is well known that there exists a family of quantum states $\Phi_C\in\mathcal{F}_s(\mathcal{H})$, parameterized by $C\in \mathcal{H}$, that behave semiclassically under the free auxiliary evolution. These are the coherent states
\begin{eqnarray*}
\Phi_C=e^{-\|C\|^2/2}\bigoplus_{n=0}^\infty
\frac{1}{\sqrt{n!}}C^{\otimes n}\,,
\end{eqnarray*}
where $C^{\otimes 0}=1\in \mathbb{C}$ and $C^{\otimes n}\in \mathcal{H}^{\otimes_s n}$ denotes the tensor product of $n$ copies of the vector $C\in \mathcal{H}$. Notice that $\quad \|\Phi_C\|=1$ irrespectively of the value of $\|C\|$. The scalar product of two coherent states $\Phi_{C_1}$ and $\Phi_{C_2}$ can be expressed in terms of the scalar product in the one-particle Hilbert space as
\begin{eqnarray*}
\langle
\Phi_{C_1}\,|\,\Phi_{C_2}\rangle=\exp\left(-\frac{1}{2}\|C_1-C_2\|^2+i\mathrm{Im}
\langle C_1,C_2\rangle\right)\,;
\end{eqnarray*}
in particular
\begin{eqnarray*}
|\langle \Phi_{C_1}\,|\,\Phi_{C_2}\rangle|=
\exp\bigg(-\frac{1}{2}\|C_1-C_2\|^2\bigg)>0\,,\quad\rm{for}\,\,\rm{all}\,\,C_1,\,\,
C_2\in \mathcal{H}\,.
\end{eqnarray*}
The inner product $\langle
\Phi_{C_1}\,|\,\Phi_{C_2}\rangle$  never vanishes but $|\langle \Phi_{C_1}\,|\,\Phi_{C_2}\rangle|$ decreases when we increase the distance between the Cauchy data
$C_1$ and $C_2$.
The class of coherent states is closed under the free dynamics defined by (\ref{free_evol}),
$$
U_0(t)\Phi_C=\exp(-itH_0)\Phi_C=\Phi_{\exp(-it\sqrt{\Theta})C}=\Phi_{C^0_t}\,.
$$
In other words, at any given time $t$ the free quantum evolution of the coherent state associated to the Cauchy data $C$ is just the coherent state associated to the classical time evolution of these Cauchy data.

\bigskip

For the full evolution the situation is, on the other hand, quite different because in this case, if $C\neq 0$,  the time evolution defined by the full physical Hamiltonian $H$ is such that
$$
U(t)\Phi_C=\exp(-itH)\Phi_C\neq \Phi_{C_t}=\exp\Big(-ite^{-h_0(C)/2} H_0\Big)\Phi_C\,.
$$
As we can see $U(t)\Phi_C$ does \textit{not} give the coherent state labeled by the classical solution $C_t$. Furthermore an argument similar to the one presented in \cite{coher} for the harmonic oscillator shows that $U(t)\Phi_C$, with $C\neq 0$, does not belong to the class of coherent states. The  case $C=0$ is special because the coherent state $\Phi_0=1\oplus 0\oplus 0\oplus \cdots\in\mathcal{F}_s(\mathcal{H})$ is both the Fock vacuum and the vacuum for the Hamiltonian $H$. It satisfies $H \Phi_0=0$ and hence $U(t)\Phi_0=\Phi_0$.

\bigskip

In the following we will give a quantitative measure of how the time evolution of the free coherent states deviates from the behavior that one would naturally demand for a \textit{bona fide} coherent state.  First, we will study the function
\begin{eqnarray*}
D_C(t)&:=&\|U_0(t)\Phi_C-U(t)\Phi_C\|^2\\
&=&2\langle\Phi_C\,|\,\big(1-\cos(t(E(H_0)-H_0))\big)\Phi_C\rangle
\\&=&2-2e^{-\|C\|^2}\sum_{n=0}^\infty\frac{1}{n!}\langle
C^{\otimes n}\,|\,\cos\big(t(E({H}_0)-{H}_0)\big)C^{\otimes n} \rangle
\end{eqnarray*}
that explicitly measures the distance between the states obtained by evolving a given coherent state with the free and the full dynamics. Second, we will consider the function
\begin{eqnarray*}
P_C(t)&:=&\langle\Phi_{C_t}\,|\,U(t)\Phi_C\rangle=  \langle \Phi_C \,|\,
\exp( it( e^{-h_0(C)/2}{H}_0-E({H}_0) ) ) \Phi_C \rangle
\\
&=&e^{-\|C\|^2}\sum_{n=0}^\infty\frac{1}{n!}\langle C^{\otimes n}\,|\,
\exp(it(e^{-h_0(C)/2}{H}_0-E({H}_0)))C^{\otimes n}\rangle\,
\end{eqnarray*}
that tells us how the full evolution of a coherent state defined by some Cauchy data deviates from the coherent state associated to the full classical evolution of the same initial data. This is done by studying the projection of one state onto the other. In particular we will consider the short and long time limits of $D_C(t)$ and $P_C(t)$. The short time limit will give us information about how fast a coherent state of the auxiliary free dynamics ceases to be semiclassical. The large time limit will lend us some information about how far from each other these states are if we let them evolve for a sufficiently long time.

\subsection{Asymptotic behavior for short times}

The behavior of these functions for short times can be obtained from the following result that can be easily derived by using a Taylor expansion.

\bigskip

\noindent Let $C\in\mathcal{D}(\Theta^{n})\subset \mathcal{H}$, then
\begin{eqnarray*}
D_C(t)&=&2\sum_{k=0}^n \frac{(-1)^{k+1}t^{2k}}{(2k)!}\langle ({H}_0-E({H}_0))^{2k}\rangle_{\Phi_C}+O(t^{2n+2})\,,\\
 P_C(t)&=&\sum_{k=0}^n \frac{(it)^k}{k!}\langle
(e^{-h_0(C)/2}{H}_0-E({H}_0))^k \rangle_{\Phi_C}+O(t^{n+1})\,,
\end{eqnarray*}
where, as usual, $\langle \mathcal{O} \rangle_\Psi=\langle \Psi |\mathcal{O}\Psi\rangle $ denotes the expectation value of the observable $\mathcal{O}$ in the normalized state $\Psi$. There are several cases that we have to analyze separately
\begin{eqnarray*}
D_C(t)&=& t^2\,\langle(E({H}_0)-{H}_0)^2\rangle_{\Phi_C} +O(t^4)\,.\\
\mathrm{Re}\Big(P_C(t)\Big)&=&1-\frac{t^2}{2}\,\langle
(e^{-h_0(C)/2}{H}_0-E({H}_0))^2\rangle_{\Phi_C} +O(t^4)\,.\\
\mathrm{Im}\Big(P_C(t)\Big)&=& t\,\langle
e^{-h_0(C)/2}{H}_0-E({H}_0)\rangle_{\Phi_C}+O(t^3)\,.
\end{eqnarray*}
First of all we see that $D_C(0)=0$ and $P_C(0)=1$. Also, as expected, the short time asymptotic behavior is controlled by the energy $E({H}_0)$. For $D_C(t)$ we see that, as long as we choose states $\Phi_C$ such that  $\langle(E({H}_0)-{H}_0)^2\rangle_{\Phi_C}$ is small the values of $D_C(t)$ will be approximately zero (they behave as a constant times $t^2$). In an analogous way, those states giving a small value for $\langle e^{-h_0(C)/2}{H}_0-E({H}_0)\rangle_{\Phi_C}$ will force $P_C(t)$ to remain close to one for a longer period of time.

\subsection{Asymptotic behavior for long times}{\label{long}}

The study of the asymptotic behavior for $t\rightarrow \infty$ is not as straightforward as the previous one and requires some work. In this case we will use the stationary phase method to obtain the sought for asymptotic behaviors. In the following it will be useful to work with finite sums instead of infinite series so, for each $N\in \mathbb{N}$, we start by defining the truncations
\begin{eqnarray*}
D_C(t,N)&:=&2-2e^{-\|C\|^2}\sum_{n=0}^N\frac{1}{n!}\langle
C^{\otimes n}\,|\,\cos\big(t(E({H}_0)-{H}_0)\big)C^{\otimes n} \rangle\,,\\
P_C(t,N)&:=&e^{-\|C\|^2}\sum_{n=0}^N\frac{1}{n!}\langle C^{\otimes
n}\,|\, \exp( it( e^{-h_0(C)/2}{H}_0-E({H}_0) ) )C^{\otimes n}\rangle\,.
\end{eqnarray*}
These functions $D_C(t,N)$ and $P_C(t,N)$ involve a finite number of terms and approximate the corresponding $D_C(t)$ and $P_C(t)$ \textit{uniformly}\footnote{The auxiliary mathematical results presented in this section are proved in the appendix.} in $t$. This means that if we fix a certain element $C\in\mathcal{H}$ and $\varepsilon>0$ there exists a natural number $N_C(\varepsilon)\in\mathbb{N}$ such that
\begin{eqnarray}
&&|D_C(t)-D_C(t,N_C(\varepsilon))|<2\varepsilon\nonumber\\
&&|P_C(t)-P_C(t,N_C(\varepsilon))|<\varepsilon\nonumber
\end{eqnarray}
irrespective of the value of $t$. This results allow us to work with the approximations given by $D_C(t,N)$ and $P_C(t,N)$.

Let us first consider the asymptotic behavior of the squared distance $D_C(t)$. If $C(r)$ is a continuous function given by the expression
$$
C(r)=\frac{1}{\sqrt{2}}\int_0^\infty c(w)J_0(wr)\mathrm{d}w\in\mathcal{H}
$$
the approximations provided by $D_C(t,N)$ have the following asymptotic behavior for $t\rightarrow\infty$ (see appendix \ref{appendix})
\begin{equation}
D_C(t,N)\sim 2-2e^{-\|C\|^2}-B_ce^{-\|C\|^2}\Gamma\left(\frac{\beta_c+1}{2}\right)
\cos\left(\frac{\pi}{4}(\beta_c+1)\right)\left(\frac{4}{t}\right)^{\frac{\beta_c+1}{2}}\,,\label{6.19}
\end{equation}
where $B_c$ and $\beta_c\geq2$ are real numbers depending on the chosen state $C(r)$. As we can see $D_C(t,N)$ approaches $2-2e^{-\|C\|^2}$ as $1/\sqrt{t^3}$ (or faster). The distance remains small when $\|C\|\rightarrow 0$ (i.e. when $C$ is close, in the $\|\cdot\|$-norm, to the value $0\in \mathcal{H}$ that labels the Fock vacuum state $\Phi_0$) and it approaches its maximum value\footnote{For unit orthogonal vectors the maximum valued of the norm of their difference is $\sqrt{2}$.} when $\|C\|\rightarrow \infty$ (due to the exponential decay in $\|C\|^2$ this limit is reached very fast).

Finally let us consider $P_C(t)$. In this case, for a continuous $C\in\mathcal{H}$, the approximations $P_C(t,N)$ have the following asymptotic behavior
$$
P_C(t,N)=e^{-\|C\|^2}+\frac{1}{\sqrt{t}}\exp(it \varrho(C))F(C,N)+O(1/t)\,,
$$
where
$$
\varrho(C):=\left(h_0(C)+2\right)e^{-h_0(C)/2}-2
$$
and $F(C,N)$ is a fixed factor that depends on $C$ and $N$. They can be obtained by applying the stationary phase method as explained in the appendix. The main conclusion that we draw from the asymptotic analysis that we have carried out in the large time limit $t\rightarrow\infty$ is that the coherent states corresponding to the free dynamics do not behave as semiclassical states for the dynamics defined by the full Hamiltonian of the system as soon as $\|C\|\sim 1$. In particular the quantum evolution $U(t)\Phi_C$ of the coherent state defined by the initial data $C$, with $ \|C\|\gg 1$, and the coherent state $\Phi_{C_t}$ labeled by the classical evolution of $C$ become almost orthogonal for large times.

\section{Quantum field operators and n-point functions}{\label{Npoint}}

The main purpose of this section is to look at the problem of quantizing Einstein-Rosen waves and related models from a perturbative perspective. This is an interesting issue because we have an exact quantization in our hands and, hence, we can compare exact results with those obtained by suitable approximations. In fact, as we have discussed elsewhere \cite{BarberoG.:2003ye, BarberoG:2004uc, BarberoG.:2004uv}, if we use asymptotic methods to extract the physical behavior of the model in terms of the relevant coupling constant (related to the Planck length) we are led to behaviors that cannot be captured by the power series expansions that one expects to get from a perturbative approach. The $n$-point functions play a very important role in quantum field theory. In fact, for the standard physical models, they are the key ingredients to construct relevant physical quantities such as the $S$-matrix. We will try here to study the structure of the $n$-point functions and compare them with those obtained from familiar QFT's such as QED. As we will see the structure of these Green functions is not the standard one corresponding to interactions defined by field-dependent potential terms. This gives us a different perspective concerning the failure of standard perturbative treatments to deal with the types of QFT's considered in this paper.

\bigskip

In the following we will use creation and annihilation operators $a^*(C)$ and $a(\overline{C})$ respectively. These are labeled with vectors $C\in\mathcal{H}$ in the one-particle Hilbert space introduced in section \ref{1phs}. They satisfy the usual commutation relations
$$
\phantom{} [a(\overline{C}_1),a^*(C_2)]=\langle C_1,C_2\rangle
\,\mathrm{Id}\,,
$$
where $\mathrm{Id}$ denotes the identity operator on  $\mathcal{F}_s(\mathcal{H})$. The conjugation $\bar{\phantom{C}}:\mathcal{H}\rightarrow
\mathcal{H}$ introduced above allows us to define subspaces of $\mathcal{H}$ consisting of purely real or imaginary vectors
$$
\mathcal{H}_R:=\{C\in \mathcal{H}\,|\,
\overline{C}=C\}\,,\quad\mathcal{H}_I:=\{C\in \mathcal{H}\,|\,
\overline{C}=-C\}\,.
$$
They are related by the equality
$$
\mathcal{H}_I=i \mathcal{H}_R\,.
$$
Now, given $f\in \mathcal{H}_R$, we can define the field and momentum operators $\phi(f)$ and $\pi(f)$ in terms of annihilation and creation operators
\begin{eqnarray*}
\phi(f)&:=&a(\overline{f}) +
a^*(f)= a(f) + a^*(f)\\
\pi(f)&:=& a(\overline{if}) + a^*(if)=-i a(f) +i
a^*(f)\,.
\end{eqnarray*}
They satisfy the commutation relations
\begin{eqnarray*}
\phantom{}[\phi(f_1),\pi(f_2)]=2i\langle
f_1,f_2\rangle\, \mathrm{Id}\,.
\end{eqnarray*}
It is possible to introduce a single operator $\Upsilon(C)$, labeled by $C=\frac{1}{2}(Q+i\Theta^{-\frac{1}{2}} P)$, to describe both the field and its canonically conjugate momentum
\begin{eqnarray*}
\Upsilon(C)&:=& a(\overline{C}) + a^*(C)\\
&=&\frac{1}{2} \Big(a(Q)+a(\overline{i\Theta^{-\frac{1}{2}}
P}) + a^*(Q)+a^*(i\Theta^{-\frac{1}{2}} P)\Big)\\
&=&\frac{1}{2}\Big(\phi(Q)+ \pi(\Theta^{-\frac{1}{2}} P)\Big)\,.
\end{eqnarray*}
The commutation relations for these operators are simply given in terms of the symplectic form (\ref{omega}) by
$$
\phantom{}[\Upsilon(C_1),\Upsilon(C_2)]=-i\omega(C_1,C_2)\, \mathrm{Id}\,.
$$

We study now the (Heisenberg image) time evolution of the $\Upsilon(C)$ from an initial instant of time $t=0$ to a generic time $t$ both under the free auxiliary dynamics and the full dynamics introduced above.

\bigskip

\noindent{\textbf{Free dynamics.}} In this case we can immediately see that
$$
\Upsilon^0(t,C):=U_0^{-1}(t)\Upsilon(C)U_0(t)=\exp(itH_0)\Upsilon(C)\exp(-itH_0)=\Upsilon(C^0_t)
$$
where $C_t^0=\exp(-it\sqrt{\Theta})C$ is the free classical evolution of the Cauchy data defined by $C$. The fact that the free dynamics can be written in such simple terms reflects in the form of the $n$-point functions $F^0_n$ defined as the vacuum expectation values of products of $\Upsilon^0(t,C)$ for different instants of time and \begin{eqnarray*}
F^0_n(t_1,C_1;t_2,C_2;\dots;t_n,C_n)&=&\langle \Upsilon^0(t_1,C_1)\Upsilon^0(t_2,C_2)\dots
\Upsilon^0(t_n,C_n) \rangle_{\Phi_0}\\
&=&\langle \Upsilon(C^0_{1\,t_1})\Upsilon(C^0_{2\,t_2})\dots
\Upsilon(C^0_{n\,t_n})\rangle_{\Phi_0}\,.
\end{eqnarray*}
In fact, it is well known that $$F^0_{2n+1}(t_1,C_1;t_2,C_2;\dots;t_{2n+1},C_{2n+1})=0$$ and $F^0_{2n}(t_1,C_1;t_2,C_2;\dots;t_{2n},C_{2n})$ can be written in terms of two-point functions
$F^0_2(t_i,C_i;t_j,C_j)$. For example the four-point function is given by
\begin{eqnarray*}
F_4^0(t_1,C_1;t_2,C_2,t_3,C_3;t_4,C_4)&=&F^0_2(t_1,C_1;t_3,C_3)F^0_2(t_2,C_2;t_4,C_4)\\
 & +&
F^0_2(t_1,C_1;t_4,C_4)F^0_2(t_2,C_2;t_3,C_3)\\&
+&F^0_2(t_1,C_1;t_2,C_2)F^0_2(t_3,C_3;t_4,C_4)\,.
\end{eqnarray*}

\bigskip

\noindent{\textbf{Full dynamics.}} Let us discuss now the evolution defined by the full Hamiltonian of the system. In this case the (Heisenberg) time evolution of the operators $\Upsilon(C)$ is given by
$$
\Upsilon(t,C):=U^{-1}(t)\Upsilon(C)U(t)=\exp(itH)\Upsilon(C)\exp(-itH)\,.
$$
The unitarity of the time evolution implies that the (equal time) commutation relations between the $\Upsilon(C)$ operators are independent of $t$,
$$
\phantom{}[\Upsilon(t,C_1),\Upsilon(t,C_2)]=-i\omega(C_1,C_2)\mathrm{Id}\,.
$$
However, at variance with the free evolution, it is clear now that
$$
\Upsilon(t,C)\neq \Upsilon(C_t)=\Upsilon\Big(\exp\big(-ite^{-h_0(\kappa^{-1}C)/2}\sqrt{\Theta}\big) C\Big)
$$
because in this case the classical dynamics $$C_t=\exp\big(-ite^{-h_0(\kappa^{-1}C)/2}\sqrt{\Theta}\big) C$$
has a non-linear depencence on the initial data $C$. This also means that it is not possible to find any Bogoliubov relation of the form
$$
U^{-1}(t)a(\overline{C})U(t)=a(\overline{A_tC})-a^*(\overline{B_t
C}), \quad\textrm{ for all } t,
$$
for any pair of operators $A_t$ and $B_t$ defined on the one-particle Hilbert space.
It is important to notice at this point that, despite the naive expectation, we have that
$$
U^{-1}(t)a(\overline{C})U(t)\neq
a(\overline{\exp(itE(\sqrt{\Theta}))C})\,.
$$
In fact the real situation is the following. The time evolution of the annihilation and creation operators $a^*(C)$ and $a(C)$ is given by expressions\footnote{These have been derived in a slightly different form in \cite{BarberoG.:2003ye}.} of the form
\begin{eqnarray}
a(t,\overline{C}):=U^{-1}(t)a(\overline{C})U(t)&=&\exp(-ite^{-H_0/2}\otimes
E(\sqrt{\Theta})^{\mathrm{right}} )
a(\overline{C})\,,\label{at1}\\
a^*(t,C):=U^{-1}(t)a^*(C)U(t)&=& a^*(C)\exp(it e^{-H_0/2} \otimes
E(\sqrt{\Theta})^{\mathrm{left}})\,.\label{at2}
\end{eqnarray}
This can be easily proved by using the identities
$$
\phantom{}[a(\overline{C}),H_0^n]=H_0^na(\overline{\Theta^\frac{n}{2}C})-H_0^na(\overline{C})\,
$$
or equivalently
$$
a(\overline{C})H_0^n=H^n_0a(\overline{\Theta^\frac{n}{2}C})\,.
$$
In the previous formulas (\ref{at1}) and (\ref{at2}) we have used a notation that tries to convey the interplay between the $C$'s that label the operators and the Hilbert space states upon which they act. Given a $n$-particle state
$V\in\mathcal{H}^{\otimes_sn}$, with
Fourier coefficients $v(w_1,\dots,w_n)$, the vector
$$
\exp(-ite^{-H_0/2}\otimes E(\sqrt{\Theta})^{\mathrm{right}} )
a(\overline{C}) V\in\mathcal{H}^{\otimes_s(n-1)}
$$
has the following Fourier coefficients
$$
\sqrt{n}\int_0^\infty \exp\Big(-ite^{-(w_1+\cdots+w_{n-1})/2}
E(w)\Big) \overline{c(w)} v(w,w_1,\dots,w_{n-1})\,\mathrm{d}w\,.
$$
This mixing, due to the interaction present in the system, introduces important complications in the computation of $n$-point functions for $n>2$ and makes it quite different from the free case.

The $n$-point functions for the Einstein-Rosen waves considered here are defined as
$$
F_n(t_1,C_1;t_2,C_2;\ldots;t_n,C_n):=\langle
\Phi_0\,|\,\Upsilon(t_1,C_1)\Upsilon(t_2,C_2)\cdots
\Upsilon(t_n,C_n)\Phi_0\rangle\,.
$$
Owing to the fact that we are dealing with an effectively interacting model these $n$-point functions behave very differently from the free case ones. The two-point function can be very easily computed in this case
\begin{eqnarray*}
F_2(t_1,C_1;t_2,C_2)&=& \langle\Phi_0\,|\,
a(t_1,\overline{C_1})a^*(t_2,C_2)\Phi_0\rangle =\langle a^*(t_1,C_1)
\Phi_0\,|\, a^*(t_2,C_2)\Phi_0\rangle
\\&=&
\langle
\exp\big(it_1E(\sqrt{\Theta})\big)C_1\,|\,\exp\big(it_2E(\sqrt{\Theta})\big)C_2
\rangle\\
&=&\int_0^\infty
\exp\Big(i(t_2-t_1)E(w)\Big)\overline{c_1(w)}c_2(w)\,\mathrm{d}w\,.
\end{eqnarray*}
These two-point functions have been studied in detail in \cite{BarberoG:2004uc,BarberoG.:2004uv}. Notice that when the coupling constant of the model\footnote{$G_3$ is the effective Newton constant per unit length in the direction of the symmetry axis.} $G=G_3\hbar$ is reintroduced ($G=G_3$ in units $\hbar=1$)  it appears in \textit{non-polynomial form} trough the expressions involving the function $E$ defined in (\ref{E}), explicitly $E(w)=\frac{1}{4G}(1-e^{-4Gw})$. This is a distinctive feature of the model that ultimately leads to behaviors that cannot be written as powers of the coupling constants.
Likewise, the four-point function can be obtained in a direct way and is given by
\begin{eqnarray*}
F_4(t_1,C_1;t_2,C_2;t_3,C_3;t_4,C_4)&=&\langle
a^*(t_2,C_2)a^*(t_1,C_1)\Phi_0\,|\, a^*(t_3,C_3)a^*(t_4,C_4)\Phi_0\rangle\\
&+&F_2(t_1,C_1;t_2,C_2)F_2(t_3,C_3;t_4,C_4)
\end{eqnarray*}
but, at variance with the situation for the free four-point function, $F_4$ cannot be written as a sum of products of two-point functions. The difference lies in what we call a \textit{mixing term}
$$
\langle a^*(t_2,C_2)a^*(t_1,C_1)\Phi_0\,|\,
a^*(t_3,C_3)a^*(t_4,C_4)\Phi_0\rangle
$$
that cannot be written as a product of two-point functions. Explicitly,
\begin{eqnarray*}
&&\langle
a^*(t_2,C_2)a^*(t_1,C_1)\Phi_0\,|\,a^*(t_3,C_3)a^*(t_4,C_4)\Phi_0\rangle=
\\
&&\phantom{\hat{a}^\dagger}=\int_0^\infty \left(\int_0^\infty
\overline{c_2(w_2)}c_4(w_2)\exp\Big(i(t_4 -t_2e^{-w_1/2})E(w_2)
+it_3e^{-w_2/2}E(w_1))\Big)\,dw_2 \right) \times
\\
&&\hspace*{4cm}\times\,\overline{c_1(w_1)}c_3(w_1) \exp\Big(-it_1E(w_1)\Big)\mathrm{d}w_1+
\\
&&\phantom{\hat{a}^\dagger \hat{a}^\dagger\,\,}+\int_0^\infty \left(\int_0^\infty
\overline{c_2(w_2)}c_3(w_2)\exp\Big(i(t_3 -t_2)e^{-w_1/2}E(w_2)\Big)\,dw_2 \right) \times
\\
&&\phantom{\hat{a}^\dagger}\hspace*{4cm} \times\overline{c_1(w_1)}c_4(w_1)
\exp\Big(i(t_4-t_1)E(w_1)\Big)
\mathrm{d}w_1\,.
\end{eqnarray*}
We will refer to this situation by saying that this last term has a mixing order of four. Notice again the non-trivial behavior of this function in terms of the coupling constant of the model.

Finally it can be shown in general that $F_{2n+1}=0$ whereas the $2n$-point functions
\begin{eqnarray*}
F_{2n}(t_1,C_1;\dots;t_{2n},C_{2n})=\hspace{11.7cm}
\\
=
\langle\Phi_0\,|\,a(t_1,\overline{C}_1)(a(t_2,\overline{C}_2)+a^*(t_2,C_2))\cdots
(a(t_{2n-1},\overline{C}_{2n-1})+a^*(t_{2n-1},C_{2n-1}))a^*(t_{2n},C_{2n})\Phi_0\rangle
\end{eqnarray*}
always have a term with maximal mixing order of $2n$. In every case there are two extreme situations as far as the mixing order of the different terms is concerned. On one hand\footnote{With the aim of simplifying some expressions we will use the following notation for products of operators $\prod_{k=1}^n{A}_k:={A}_1{A}_2\cdots {A}_n\,.$} we can have
\begin{eqnarray*}
\langle\Phi_0\,|\prod_{k=1}^{n}\,a(t_{2k-1},\overline{C}_{2k-1})a^*(t_{2k},C_{2k})\Phi_0\rangle= \prod_{k=1}^nF_2(t_{2k-1},C_{2k-1};t_{2k},C_{2k})
\end{eqnarray*}
that can be written as a product of two-point functions.
On the other hand
\begin{eqnarray*}
\langle\Phi_0\,|\prod_{k=1}^{n}\,a(t_k,\overline{C_k})\prod_{s=n+1}^{2n}a^*(t_s,C_{s})\Phi_0\rangle=
\langle
\prod_{k=1}^{n}\,a^*(t_k,C_k)\Phi_0\,|\prod_{s=n+1}^{2n}a^*(t_s,C_{s})\Phi_0\rangle
\end{eqnarray*}
is maximally mixed (in fact the mixing order is higher than the maximum present for $2(n-1)$-point functions).

\section{Conclusions}

We have studied the quantization of Einstein-Rosen waves in the reduced phase space obtained by imposing the asymptotic flatness condition of \cite{Ashtekar:1994ds} and using an asymptotic, unit, timelike,  Killing vector field to parameterize the time evolution. We have discussed two different types of issues that are relevant to understand quantum Einstein-Rosen waves. The first issue that we have considered is related to the semiclassical limit of the system. Specifically we have studied to what extent the coherent states corresponding to the free auxiliary dynamics of the model can be thought of as semiclassical states under the evolution defined by the full non-quadratic Hamiltonian. The conclusion that we draw from our analysis is that for short periods of time (with a length determined by the Hamiltonian $H_0$ as expected on general grounds) the free coherent states can, indeed, be considered as semiclassical. In the long time limit we recover \textit{from a dynamical point of view} the results of Ashtekar \cite{Ashtekar:1996yk} about the existence of large quantum effects in the system. Specifically we see that the free coherent states $\Phi_C$ with low occupation number ($\|C\|\sim 1$) do not behave semiclassically and it gets worse and worse for larger values of $\|C\|$. We have discussed this by studying the distance and the mutual projections of certain states obtained by considering the different time evolutions relevant in the model (the auxiliary one given by $H_0$ and the full physical one). If we consider the squared distance, as a function of time, between the states obtained by evolving coherent states with the auxiliary evolution and the full evolution we see that it approaches the value $2-2\exp(-\|C\|^2)$. If $\|C\|$ is very small this distance remains small whereas it becomes significant once $\|C\|\sim 1$. Finally for larger values of $\|C\|$ the distance approaches the maximum value for orthogonal states. Similar conclusions can be reached by studying suitable projections. We want to emphasize that we have not proved the impossibility of finding good semiclassical states for the model but only that not all the semiclassical states for the free auxiliary model can be considered as such for the full dynamics. An interesting open problem is to find a sufficiently large class of semiclassical states representing classical ER waves corresponding to arbitrary Cauchy data $C$.

A second point that we have studied is the mathematical structure of the $n$-point functions. The main reason to do this is to get some information about the possible perturbative analysis of the system. We have seen that the structure of the $n$-point functions is different from the one corresponding to a free QFT. This is noteworthy because the formalism that we have used here relies on the fact that our model can be conveniently described in terms of an auxiliary free model. The structure of the $n$-point functions, for which we are able to give closed form expressions, displays the distinctive features of an interacting model because they cannot be written simply in terms of two-point functions. Also the type of non-local interaction underlying the model shows up in the detailed form of these objects that differ from the ones obtained for familiar systems where the interaction is just given by a field-dependent potential. This is compatible with the known fact (discussed elsewhere) that the asymptotic approach to the study of physical observables for this model leads to expansions in terms of the relevant coupling constant (that can be interpreted as an effective Planck length) that are incompatible with any power series \cite{BarberoG.:2004uv,BarberoG.:2005ge}.

We want to conclude by remarking that the present model can be exactly solved. In fact the exact evolution operator and their matrix elements can be exactly written. This means, in particular, that there is no need to separately consider the $n$-point functions to construct physical objects such as the $S$-matrix. Of course $n$-point functions are interesting objects with important physical interpretations (see \cite{BarberoG.:2003ye, FernandoBarbero:2006gd}) so it makes sense to understand how they can be obtained as we have done here.

\begin{acknowledgments}
The authors want to thank Daniel G\'omez Vergel for his comments and careful reading of the manuscript. I\~{n}aki Garay acknowledges the financial support provided by the Spanish
Ministry of Science and Education (MEC) under the FPU program. This work
is also supported by the Spanish MEC under the research grant
FIS2005-05736-C03-02.

\end{acknowledgments}

\appendix

\section{Asymptotic expansions}{\label{appendix}}

This appendix contains the proofs of several results used to obtain the asymptotic behavior of $D_C(t)$ and $P_C(t)$ in the large $t$ asymptotic limit studied in subsection \ref{long}. We will write them in the form of propositions.

\bigskip

\begin{prop}
Given $C\in\mathcal{H}$ and
$\varepsilon>0$ there is a number $N_C(\varepsilon)\in \mathbb{N}$, independent of $t\in\mathbb{R}$, such that
\begin{eqnarray*}
|D_C(t)-D_C(t,N_C(\varepsilon))|<2\varepsilon
\end{eqnarray*}
and
$$ |P_C(t)-P_C(t,N_C(\varepsilon))|<\varepsilon\,.$$
\end{prop}

\bigskip

The proof is based on the fact that it is always possible to find $N_C(\varepsilon)\in \mathbb{N}$ in such a way that
$$
e^{-\|C\|^2}\sum_{n=N_C(\varepsilon)}^\infty\frac{1}{n!}\|
C\|^{2n}<\varepsilon\,;
$$
and hence,
\begin{eqnarray*}
|D_C(t)-D_C(t,N_C(\varepsilon))|&=&\bigg|2e^{-\|C\|^2}\sum_{n=N_C(\varepsilon)}^\infty\frac{1}{n!}\langle
C^{\otimes n}\,|\,\cos\big(t(E({H}_0)-{H}_0)\big)C^{\otimes n}
\rangle\bigg|\\
&\leq&2e^{-\|C\|^2}\sum_{n=N_C(\varepsilon)}^\infty\frac{1}{n!}\bigg|\langle
C^{\otimes n}\,|\,\cos\big(t(E({H}_0)-{H}_0)\big)C^{\otimes n}
\rangle\bigg|\\
&\leq&2e^{-\|C\|^2}\sum_{n=N_C(\varepsilon)}^\infty\frac{1}{n!}\|
C^{\otimes n}\|\cdot\|\cos\big(t(E({H}_0)-{H}_0)\big)C^{\otimes n}\|\\
&\leq& 2e^{-\|C\|^2}\sum_{n=N_C(\varepsilon)}^\infty\frac{1}{n!}\|
C^{\otimes n}\|^2=2e^{-\|C\|^2}\!\!\!\!\sum_{n=N_C(\varepsilon)}^\infty\frac{1}{n!}\|
C\|^{2n}<2\varepsilon.\,
\end{eqnarray*}
We also have
\begin{eqnarray*}
|P_C(t)-P_C(t,N_C(\varepsilon))|&=&
\bigg|e^{-\|C\|^2}\sum_{n=N_C(\varepsilon)}^\infty\frac{1}{n!}\langle
C^{\otimes n}\,|\, \exp(it(e^{-H_0(C)/2}{H}_0-E({H}_0)))C^{\otimes
n}\rangle\bigg|\\&\leq&
e^{-\|C\|^2}\sum_{n=N_C(\varepsilon)}^\infty\frac{1}{n!}\bigg|\langle
C^{\otimes n}\,|\, \exp(it(e^{-H_0(C)/2}{H}_0-E({H}_0)))C^{\otimes
n}\rangle\bigg|
\\&\leq& e^{-\|C\|^2}\sum_{n=N_C(\varepsilon)}^\infty\frac{1}{n!}\|
C^{\otimes n}\|\cdot \|C^{\otimes
n}\|=e^{-\|C\|^2}\!\!\!\!\sum_{n=N_C(\varepsilon)}^\infty\frac{1}{n!}\|
C\|^{2n}<\varepsilon\,.
\end{eqnarray*}
This result allows us to work with the truncations given by $D_C(t,N)$ and $P_C(t,N)$.

\bigskip

\begin{prop} Let
$$
C(r)=\frac{1}{\sqrt{2}}\int_0^\infty c(w)J_0(w r)\mathrm{d} w\,\in\mathcal{H}\,,
$$
and let us assume  $|c(w)|^2=B_cw^{\beta_c}+O(w^{\beta_c+2})\quad(w\rightarrow 0^+)$, with $B_c\neq0$ and $\beta_c\geq 2$ real constants depending on the state $C$. Then the approximations $D_C(t,N)$ satisfy, in the limit $t\rightarrow \infty$,
\begin{eqnarray*}
D_C(t,N)\sim 2-2e^{-\|C\|^2}-B_ce^{-\|C\|^2}\Gamma\left(\frac{\beta_c+1}{2}\right)
\cos\left(\frac{\pi}{4}(\beta_c+1)\right)\left(\frac{4}{t}\right)^{\frac{\beta_c+1}{2}}\,.
\end{eqnarray*}
As a consequence, $D_C(t,N)$ approaches $2-2e^{-\|C\|^2}$ as $(1/t)^{3/2}$ or faster.
\end{prop}

\bigskip

This result follows from
\begin{eqnarray*}
D_C(t,N)=2-2e^{-\|C\|^2}\sum_{n=0}^N\frac{1}{n!}A^{\scriptscriptstyle C}_n \,,
\end{eqnarray*}
where $A^{\scriptscriptstyle C}_0:=1$ and
\begin{eqnarray*}
&&A^{\scriptscriptstyle C}_n:=\langle C^{\otimes n}\,|\,\cos\big(t(E({H}_0)-{H}_0)\big)C^{\otimes n}
\rangle=\\
&&\phantom{A^C_n:}=\int_{[0,\infty)^n} \cos\bigg(t\big(\sum_{j=1}^nw_j
-E(\sum_{j=1}^nw_j)\big)\bigg)|c(w_1)|^2\cdots
|c(w_n)|^2\,\mathrm{d}w_1\cdots \mathrm{d}w_n\\
&&\phantom{A^C_n:}=\frac{1}{2}\int_0^\infty \!\!\exp\big(it\big(z
-E(z)\big)\big)G_n(z,C)\,\mathrm{d}z+\frac{1}{2}\int_0^\infty
\!\!\exp\big(-it\big(z -E(z)\big)\big)G_n(z,C)\,\mathrm{d}z\,.
\end{eqnarray*}
In the last integral we have performed the following change of variables
$$
(w_1,\dots,w_{n-1},w_n)\mapsto (w_1,\dots,w_{n-1},z)\,,\, \textrm{ where } z=\sum_{j=1}^nw_j\,,
$$
and defined the functions $G_n(z,C)$ as
\begin{eqnarray*}
G_n(z,C):=\int_{\prod_{k=1}^{n-1}[0,z-\sum_{j=1}^{k-1}w_j)}
|c(z-\sum_{j=1}^{n-1}w_j)|^2\prod_{k=1}^{n-1} |c(w_k)|^2
\mathrm{d}w_1\cdots \mathrm{d}w_{n-1}\,.
\end{eqnarray*}
These behave, in the $z\rightarrow0$ limit,  as $G_n(z,C)\sim B_c^nz^{n(\beta_c+1)-1}$  with $B_c$ and $\beta_c\geq 2$ real constants that depend on the state that we have chosen to start with (notice that, as we pointed out in subsection  \ref{modes}, the Fourier coefficients $c(w)$ satisfy $c(0)=0$). Hence, the stationary phase method \cite{bender} gives the following asymptotics for $A^{\scriptscriptstyle C}_n$, $n\geq1$,  in the $t\rightarrow\infty$ limit
$$
A^{\scriptscriptstyle C}_n(t)\sim \frac{B_c^n }{2}\,\Gamma\left(\frac{n(\beta_c+1)}{2}\right) \cos\left(\frac{\pi}{4}(n(\beta_c+1))\right)\left(\frac{4}{t}\right)^{\frac{n(\beta_c+1)}{2}}\,.
$$
We conclude that the main contribution to the asymptotic expansion of $D_C(t,N)$ is given by $2-2e^{-\|C\|^2}(A^{\scriptscriptstyle C}_0+A^{\scriptscriptstyle C}_1)=2-2e^{-\|C\|^2}(1+A^{\scriptscriptstyle C}_1)$. This way we finally get (\ref{6.19}).

\bigskip

\begin{prop} If  $C\in\mathcal{H}$ (regular enough) then $P_C(t,N)$ satisfies the identity
$$
P_C(t,N)=e^{-\|C\|^2}+\frac{\exp\big(it\varrho(C)\big)}{\sqrt{t}} F(C,N)
+O(1/t)\quad (t\rightarrow \infty)\,,
$$
where
$$
\varrho(C):=(h_0(C)+2)e^{-h_0(C)/2}-2
$$
and $F(C,N)$ is a function that depends only on $C$ and $N\in\mathbb{N}$.
\end{prop}

\bigskip

The proof is a straightforward application of the stationary phase method \cite{bender} to
\begin{eqnarray*}
&&\hspace{-0.7cm}\langle C^{\otimes n}\,|\,
\exp(it(e^{-h_0(C)/2}{H}_0-E({H}_0)))C^{\otimes n}\rangle= \int_0^\infty
\exp\bigg(it\big(e^{-h_0(C)/2}z-E(z)\big)\bigg)\, G_n(z,C)\,
\mathrm{d}z\\
&&\hspace{0.5cm}=\sqrt{4\pi}e^{h_0(C)/4}G_n(h_0(C),C)\frac{\exp\big(it\varrho(C)+\frac{i\pi}{4}\big)}
{\sqrt{t}}+O(t^{-1})\,,\quad n\geq 1\,.
\end{eqnarray*}
Notice that
$$
\langle C^{\otimes 0}\,|\,
\exp(it(e^{-h_0(C)/2}{H}_0-E({H}_0)))C^{\otimes 0}\rangle=1\,.
$$

\end{document}